\documentclass{vldb}

\usepackage{latexsym}
\usepackage{amsmath}
\usepackage{amssymb}
\usepackage{pst-tree}

\def\punto{$\hspace*{\fill}\Box$}
\newcommand{\nop}[1]{}
\newcommand{\tuple}[1]{{\langle#1\rangle}}

\newtheorem{theorem}{Theorem}[section]

\newtheorem{example}[theorem]{Example}
\newtheorem{algorithm}[theorem]{Algorithm}
\newtheorem{definition}[theorem]{Definition}
\newtheorem{proposition}[theorem]{Proposition}

\newtheorem{corollary}[theorem]{Corollary}
\newtheorem{lemma}[theorem]{Lemma}
\newtheorem{remark}[theorem]{Remark}

\title{Conditioning Probabilistic Databases}

\numberofauthors{2}

\author{
\alignauthor Christoph Koch\\
       \affaddr{Department of Computer Science} \\
       \affaddr{Cornell University} \\
       \affaddr{Ithaca, NY 14853, USA} \\
       \affaddr{koch@cs.cornell.edu}
\alignauthor Dan Olteanu \\
       \affaddr{Computing Laboratory}\\
       \affaddr{Oxford University} \\
       \affaddr{Oxford, OX1 3QD, UK} \\
       \affaddr{dan.olteanu@comlab.ox.ac.uk}
}

\toappear{}

\begin{document}
\maketitle

\begin{abstract}
Past research on probabilistic databases has studied the problem
of answering queries on a static database. Application scenarios of
probabilistic databases however often involve the conditioning of a
database using additional information in the form of new evidence. The
conditioning problem is thus to transform a probabilistic database of
priors into a posterior probabilistic database which is materialized
for subsequent query processing or further refinement.
It turns out that the conditioning problem is closely related to
the problem of computing exact tuple confidence values.

It is known that exact confidence computation is an NP-hard problem.
This has led researchers to consider approximation techniques for
confidence computation.  However, neither conditioning nor exact
confidence computation can be solved using such techniques.
In this paper we present efficient techniques for both problems.
We study several problem decomposition methods and heuristics
that are based on the most successful search techniques from
constraint satisfaction, such as the
Davis-Putnam algorithm. We complement this with a thorough
experimental evaluation of the algorithms proposed.
Our experiments show that our exact algorithms scale well to realistic database
sizes and can in some scenarios compete with the most efficient
previous approximation algorithms.
\end{abstract}

\section{Introduction}
\label{sec:introduction}

Queries on probabilistic databases have numerous applications at the
interface of databases and information retrieval \cite{FR1997}, data
cleaning \cite{AKO07WSD}, sensor data, tracking moving objects, crime
fighting \cite{BDSHW2006}, and computational science
\cite{dalvi07efficient}.

A core operation of queries on probabilistic databases is the computation of confidence
values of tuples in the result of a query. In short, the confidence in a tuple $t$ being
in the result of a query on a probabilistic database is the combined probability weight
of all possible worlds in which $t$ is in the result of the query.

By extending the power of query languages for probabilistic data\-bases,
new applications beyond the mere retrieval of tuples and their confidence become possible.
An essential operation that allows for new applications is {\em conditioning}\/,
the operation of removing possible worlds which do not satisfy a given condition from a
probabilistic database. Subsequent query operations will apply to the reduced data\-base,
and a confidence computation will return {\em conditional probabilities}\/ in the
Bayesian sense with respect to the original database.
Computing conditioned probabilistic data\-ba\-ses has natural and important applications in
virtually all areas in which probabilistic databases are useful. For example, in data cleaning,
it is only natural to start with an uncertain database and then clean it -- reduce uncertainty --
by adding constraints or additional information. More generally, conditioning allows us to
start with a database of prior probabilities, to add in some {\em evidence}\/, and take it to
a posterior probabilistic database that takes the evidence into account.

Consider the example of a probabilistic database of social security numbers
(SSN) and names of individuals extracted
from paper forms using OCR software. If
a symbol or word cannot be clearly identified, this software will
offer a number of weighted alternatives.
The database
\begin{center}
\begin{tabular}{l|cc}
R & SSN &             NAME \\
\hline
& \{ 1 (p=.2) $\mid$ 7 (p=.8) \} & John \\
& \{ 4 (p=.3) $\mid$ 7 (p=.7) \} & Bill \\
\end{tabular}
\end{center}
represents four possible worlds (shown in Figure~\ref{fig:john_bill_worlds}),
modelling that John has either
SSN 1 or 7, with probability .2 and .8 (the paper form may contain
a hand-written symbol that can either be read as a European ``1''
or an American ``7''), respectively,
and Bill has either SSN 4 or 7, with probability .3 and .7, respectively.
We assume independence between John's and Bill's alternatives, thus the world in which
John has SSN 1 and Bill has SSN 7 has probability $.2 \cdot .7 = .14$.

If $A_x$ denotes the event that Bill has SSN $x$, then
$P(A_4) = .3$ and $P(A_7) = .7$.
We can compute these probabilities in a probabilistic database by asking
for the confidence values of  the tuples in the result of the query
\[
\mbox{select SSN, conf(SSN) from R where NAME = 'Bill';}
\]
which will result in the table
\begin{center}
\begin{tabular}{l|cc}
Q & SSN & CONF \\
\hline
& 4 & .3 \\
& 7 & .7 \\
\end{tabular}
\end{center}

\begin{figure}
\begin{center}
\begin{tabular}{cc}
\begin{tabular}{l|cc}
$R^1$ & SSN & NAME \\
\hline
     & 1 & John \\
     & 4 & Bill \\
\hline
\end{tabular}
&
\begin{tabular}{l|cc}
$R^2$ & SSN & NAME \\
\hline
     & 7 & John \\
     & 4 & Bill \\
\hline
\end{tabular}
\\
P = .06
&
P = .24
\\[2ex]
\begin{tabular}{l|cc}
$R^3$ & SSN & NAME \\
\hline
     & 1 & John \\
     & 7 & Bill \\
\hline
\end{tabular}
&
\begin{tabular}{l|cc}
$R^4$ & SSN & NAME \\
\hline
     & 7 & John \\
     & 7 & Bill \\
\hline
\end{tabular}
\\
P = .14
&
P = .56
\end{tabular}
\end{center}

\vspace{-5mm}

\caption{The four worlds of the input database.}
\label{fig:john_bill_worlds}
\end{figure}

Now suppose we want to use the additional knowledge that social
security numbers are unique.
We can express this using a functional dependency
SSN $\rightarrow$ NAME. Asserting this constraint, or conditioning
the probabilistic database using the constraint, means to
eliminate all those worlds in which the functional dependency does not hold.

Let $B$ be the event that the functional dependency holds.
Conceptually, the database conditioned with $B$ is obtained by removing world $R^4$ (in which John and Bill
have the same SSN) and renormalizing the probabilities of the remaining worlds to have them again sum up to 1, in this
case by dividing by $.06+.24+.14 = .44$. We will think of conditioning as an operation
assert[$B$] that reduces uncertainty by declaring worlds in which $B$ does not hold impossible.
 
Computing tuple confidences for the above query
on the original database will give us, for each possible SSN value $x$ for
Bill, the
probabilities $P(A_x)$, while on the database conditioned with $B$ it
will give a table of social security numbers $x$ and
conditional probabilities $P(A_x \mid B)$.
For example, the conditional probability of Bill having SSN 4
given that social security numbers are unique is
\[
P(A_4 \mid B) = \frac{P(A_4 \land B)}{P(B)} = \frac{.3}{.44} \approx .68.
\]

Using this definition, we could alternatively have computed the
conditional probabilities by combining the results of two confidence
computations,
\begin{tabbing}
~~~~~~ \=
   select SSN, P1/P2 \\
\> from \=(\=select SSN, conf(SSN) P1 from R, B \\
\>      \> \>where NAME = 'Bill'), \\
\>      \>(select conf() P2 from B);
\end{tabbing}
where $B$ is a Boolean query that is true if the functional dependency holds
on $R$.

Unfortunately, both conditioning and confidence computation are NP-hard
problems. Nevertheless, their study is justified by their obvious relevance
and applications.
While conditioning has not been previously studied in the context of
probabilistic databases,
previous work on confidence computation has aimed at
cases that admit polynomial-time query evaluation and
at approximating confidence values  \cite{dalvi07efficient}.

Previous work often assumes that confidence values
are computed at the end of a query, closing the possible worlds semantics of the
probabilistic database and returning a complete, nonprobabilistic relation of tuples
with numerical confidence values that can be used for decision making. In such a context,
techniques that return a reasonable approximation of confidence values may be acceptable.

In other scenarios we do not want to accept approximate confidence values
because errors made while computing these estimates aggregate and grow,
causing users to make wrong decisions based on the query results.
This is particularly true in compositional query languages for probabilistic
databases,
where confidence values computed in a subquery form part
of an intermediate result that can be
accessed and used for filtering the data in subsequent query operations \cite{Koch2008}.

Similar issues arise when confidence values can be inserted into the
probabilistic database through updates and may be used in subsequent queries.
For example, data cleaning is a scenario where we, on one hand, want to
materialize the result of a data transformation in the database once and
for all (rather than having to redo the cleaning steps every time a query is
asked) and on the other hand do not want to store incorrect probabilities that
may affect a very large number of subsequent queries.
Here we need techniques for conditioning and
exactly computing confidence values.

\nop{
To the best of our knowledge, the efficient computation of queries that define conditional
probabilities in the context of probabilistic databases has not been
studied previously.
A single conditional probability can be computed as a ratio
of two confidence values, by the previous formula. This however requires
that the system supports rather powerful queries. For instance, the previous example query is clearly
of a form for which no efficient query processing techniques are known.
} 

Exact confidence computation is particularly
important in queries in which confidence
values are used in comparison predicates. For an example, let us add a third
person, Fred, to the database whose SSN is either 1 or 4, with equal
probability. If we again condition using the
functional dependency SSN $\rightarrow NAME$, we have only two possible
worlds, one in which John, Bill, and Fred
have social security numbers 1, 7, and 4, respectively, and one in which their SSN are 7, 4, and 1.
If we now ask for the social security numbers that are in the database for
certain,
\[
\mbox{select SSN from R where conf(SSN) = 1;}
\]
we should get three tuples in the result.  Monte Carlo simulation
based approximation algorithms will do very badly on such queries.
Confidence approximation using a
Karp-Luby-style algorithm \cite{KL1983,dalvi07efficient, RDS07}
will independently underestimate each tuple's confidence
with probability $\approx$.5. Thus the probability that at least one tuple is
missing from the result of such a query is very high (see also \cite{Koch2008}.

In this paper, we develop efficient algorithms for computing exact
confidences and for conditioning
probabilistic databases. The detailed contributions are as follows.
\begin{itemize}
\addtolength{\topsep}{-0.3ex}
\addtolength{\labelsep}{-0.3ex}
\addtolength{\itemsep}{-1ex}
\item
In most previous models of probabilistic data\-bases over finite
world-sets, computing tuple confidence values essentially means the
weighted counting of solutions to constraint structures closely
related to disjunctive normal form formulas.  Our notion of such
structures are the world-set descriptor sets, or {\em ws-sets} for
short. We formally introduce a probabilistic database model that is
known to cleanly and directly generalize many previously considered
probabilistic database models (cf.\ \cite{AJKO07}) including, among
others, various forms of tuple-independence models
\cite{dalvi07efficient,miller06clean}, ULDBs \cite{BDSHW2006},
product decomposition \cite{AKO07WSD}, and c-table-based models \cite{AJKO07}.
We use this framework to study
exact confidence computation and conditioning. The results
obtained are thus of immediate relevance to all these models.

\item
We study properties of ws-sets that are essential to relational algebra query evaluation
and to the design of algorithms for the two main problems of the paper.

\item
We exhibit the fundamental, close relationship between the two problems.

\item
We develop {\em ws-trees}\/, which capture notions of structural decomposition
of ws-sets based on probabilistic independence and world-set disjointness.
Once a ws-tree has been obtained for a given
ws-set, both exact confidence computation and conditioning are feasible in linear time.
The main problem is thus to efficiently find small ws-tree decompositions.

\item
To this end, we develop a decomposition procedure motivated by the
Davis-Putnam (DP) procedure for checking Propositional Satisfiability
\cite{DP1960}. DP, while many decades old, is still the basis of the
best exact solution techniques for the NP-complete Satisfiability
problem.  We introduce two decomposition rules, variable elimination
(the main rule of DP) and a new independence decomposition rule, and
develop heuristics for chosing among the rules.

\item
We develop a database conditioning algorithm based on ws-tree decompositions
and prove its correctness.

\item
We study ws-set simplification and elimination techniques that can be either used as an
alternative to the DP-based procedure or combined with it.

\item
We provide a thorough experimental evaluation of the algorithms presented
in this paper. We also experimentally compare our exact techniques for
confidence computation with
approximation based on Monte Carlo simulation. 
\end{itemize}

The structure of the paper follows the list of contributions.


\section{Probabilistic Databases}
\label{sec:pdb}

\begin{figure}[t]
\[
\begin{tabular}{l|lll}
$W$ & Var & Dom & P \\
\hline
    & $j$ & 1  & .2   \\
    & $j$ & 7  & .8   \\
    & $b$ & 4  & .3   \\
    & $b$ & 7  & .7   \\
\end{tabular}
\hspace{.5cm}
\begin{tabular}{l|c|cc}
$U_R$ & WSD & SSN & NAME \\
\hline
  & $\{ j \mapsto 1 \}$ & 1 & John \\
  & $\{ j \mapsto 7 \}$ & 7 & John \\
  & $\{ b \mapsto 4 \}$ & 4 & Bill \\
  & $\{ b \mapsto 7 \}$ & 7 & Bill \\
\end{tabular}
\]
 \centering

\vspace{-5mm}

\caption{Probabilistic database with ws-descriptors made explicit
 and defined by world-table $W$.}
\label{fig:pdb} 
\end{figure}

We define \textit{sets of possible worlds} 
following U-relational databases~\cite{AJKO07}.
Consider a finite set of independent random variables
ranging over finite domains.
Probability distributions over the possible worlds are
defined by assigning a probability
$P(\{ x \mapsto i \})$ to each assignment of
a variable $x$ to a constant of its domain, $i \in \mbox{Dom}_x$, such that the
probabilities of all assignments of a given variable sum up to one.
We represent the set of variables, their
domains, and probability distributions relationally by a \textit{world-table}
$W$ consisting of all triples $(x,i,p)$ of
variables $x$, values $i$ in the domain of $x$, and the associated
probabilities $p = P(\{ x \mapsto i \})$.

A \textit{world-set descriptor} is a set of assignments $x \mapsto i$
with $i \in \mbox{Dom}_x$ that is functional, i.e.\ a partial function from variables
to domain values.
If such a world-set descriptor $d$ is a {\em total}\/ function, then it identifies
a possible world. Otherwise, it
denotes all those possible
worlds $\omega(d)$ identified by total functions $f$ that can be obtained
by extension of $d$. (That is, for all $x$ on which $d$
is defined, $d(x)=f(x)$.)  
Because of the independence of the variables,
the aggregate probability of these worlds is
\[
P(d) = \prod_{\{ x \mapsto i \} \subseteq d} P(\{ x \mapsto i \}).
\]
If $d = \emptyset$, then $d$ denotes the set of all possible worlds.

We say that two ws-descriptors $d_1$ and $d_2$ are {\em consistent}\/ iff
their union (as sets of assignments) is functional.


A {\em ws-set} is a set of ws-descriptors $S$ and
represents the world-set computed as the union of the world-sets
represented by the ws-descriptors in the set.  We define the semantics
of ws-sets using the (herewith overloaded) function $\omega$ extended to
ws-sets,
$\omega(S) := \underset{d\in S}{\bigcup} (\omega(d))$.


A \textit{U-relation} over schema $\Sigma$ and world-table
$W$ is a set of tuples over $\Sigma$, where we associate to each tuple
a ws-descriptor over $W$. A \textit{probabilistic database} over
schema $\{\Sigma_1,\ldots,\Sigma_n\}$ and world-table $W$ is a set of
$n$ U-relations, each over one schema $\Sigma_i$ and $W$.
A probabilistic database represents a set of databases, one database
for each possible world defined by $W$. To obtain a possible world in the
represented set, we first choose a total valuation $f$ over $W$. We
then process each probabilistic relation $R_i$ tuple by tuple. If $f$
extends the ws-descriptor $d$ of
a tuple $t$, then $t$ is in the relation $R_i$ of that database.

\begin{example}
\label{ex:running-wtable}
\em
Consider again the probabilistic database of social security numbers and
names given in Figure~\ref{fig:john_bill_worlds}.
Its representation in our formalism is given in Figure~\ref{fig:pdb}.
The world-table $W$ of Figure~\ref{fig:pdb} defines two variables
$j$ and $b$ modeling the social security numbers of John and Bill, with domains $\{1,7\}$ and $\{4,7\}$ respectively.
The probability
of the world defined by $f = \{j\mapsto 7,b\mapsto
7\}$ is $.8\cdot .7=.56$. The total valuation $f$ extends the ws-descriptors of the
second and fourth tuple of relation $U_R$, thus the relation $R$ in world $f$ is
\{ (7, John), (7, Bill) \}. \punto
\end{example}

\begin{remark}\em
Leaving aside the probability distributions of the variables which are
represented by the $W$ table, U-relations are essentially restricted
c-tables \cite{IL1984} in which the global condition is ``true'',
variables must not occur in the tuples, and each local condition must
be a conjunction of conditions of the form $x=v$ where x is a variable
and $v$ is a constant. Nevertheless, it is known that U-relations are
a complete representation system for probabilistic databases over
nonempty finite sets of possible worlds.

U-relations can be used to represent attribute-level uncertainty using
vertically decomposed relations.
For details on this, we refer to \cite{AJKO07}. All results in
this paper work in the context of attribute-level uncertainty.

The efficient execution of the operations of positive relational algebra on
such databases was described in that paper as well. Briefly, if U-relations
$U_R$ and $U_S$
represent relations $R$ and $S$, then selections $\sigma_{\phi} R$ and
projections $\pi_{\overline{A}} R$ simply translate into
$\sigma_{\phi} U_R$ and $\pi_{WSD, \overline{A}} U_R$,
respectively.
Joins $R \bowtie_\phi S$ translate into
$U_R \bowtie_{\phi \land \psi} U_S$ where $\psi$ is the condition that the
ws-descriptors of the two tuples compared are consistent with each other
(i.e., have a common extension into a total valuation).
The set operations easily follow from the analogous operations on ws-sets that
will be described below, in Section~\ref{sec:setop}. 
\punto
\end{remark}

\begin{example}\em
\label{ex:fd}
The functional dependency SSN $\rightarrow$ NAME on the probabilistic
database of Figure~\ref{fig:pdb} can be expressed as a boolean
relational algebra query as the complement of $\pi_\emptyset(R
\bowtie_\phi R)$ where $\phi := (1.SSN=2.SSN \land 1.NAME \neq
2.NAME)$.  We turn this into the query
\[
\pi_{WSD}(U_R \bowtie_{\phi \land
1.WSD \;\mathrm{consistent \ with}\; 2.WSD} U_R).
\] 
over our representation, which results in the ws-set
$\{ j \mapsto 7, b \mapsto 7 \}$.
The complement of this with the world-set given by the $W$ relation,
$\{ \{ j \mapsto 1 \}, \{ j \mapsto 7 \},
    \{ b \mapsto 4 \}, \{ b \mapsto 7 \} \}$, is
$\{ \{ j \mapsto 1 \}, \{ j \mapsto 7, b \mapsto 4 \} \}$.
(Note that this is just one among a set of equivalent solutions.)
\punto
\end{example}

\section{Properties of ws-descriptors}
\label{sec:properties}

In this section we investigate properties of ws-descriptors and show
how they can be used to efficiently implement various set operations
on world-sets without having to enumerate the worlds. This is
important, for such sets can be extremely large in practice:
\cite{AKO07WSD,AJKO07} report on experiments with $10^{10^6}$ worlds.

\subsection{Mutex, Independence, and Containment}

Two ws-descriptors $d_1$ and $d_2$ are (1)
\textit{mutually exclusive} (mutex for short) if they represent
disjunct world-sets, i.e., $\omega(d_1)\cap\omega(d_2)=\emptyset$, and
(2) \textit{independent} if there is no valuation of the variables in
one of the ws-descriptors that restricts the set of possible
valuations of the variables in the other ws-descriptor (that is, $d_1$
and $d_2$ are defined only on disjoint sets of variables). A
ws-descriptor $d_1$ is \textit{contained} in $d_2$ if the world-set of
$d_1$ is contained in the world-set of $d_2$, i.e.,
$\omega(d_1)\subseteq\omega(d_2)$. \textit{Equivalence} is mutual
containment.

Although ws-descriptors represent very succinctly possibly very large
world-sets, all aforementioned properties can be efficiently checked
at the syntactical level: $d_1$ and $d_2$, where all variables with
singleton domains are eliminated, are (1) mutex if there is a variable
with a different assignment in each of them, and (2) independent if
they have no variables in common; $d_1$ is contained in $d_2$ if $d_1$
extends $d_2$.

\begin{example}\label{ex:properties}\em
Consider the world-table of Figure~\ref{fig:pdb} and the
ws-descriptors $d_1=\{ j \mapsto 1 \}$, $d_2=\{j \mapsto 7\}$,
$d_3=\{ j \mapsto 1, b \mapsto 4\}$, and $d_4=\{ b \mapsto 4\}$. Then,
the pairs $(d_1, d_2)$ and $(d_2, d_3)$ are mutex, $d_3$ is contained
in $d_1$, and the pairs $(d_1,d_4)$ and $(d_2,d_4)$ are
independent.\punto
\end{example}

We also consider the mutex, independence, and equivalence properties
for ws-sets. Two ws-sets $S_1$ and $S_2$ are mutex (independent)
iff $d_1$ and $d_2$ are mutex
(independent) for any $d_1 \in S_1$ and $d_2 \in S_2$.
Two ws-sets are equivalent if they represent the same world-set.


\begin{example}\em
We continue Example~\ref{ex:properties}. The ws-set $\{d_1\}$ is mutex
with $\{d_2\}$. $\{d_1,d_2\}$ is independent from $\{d_4\}$.  At a
first glance, it looks like $\{d_1,d_2\}$ and $\{d_3,d_4\}$ are
neither mutex nor independent, because $d_1$ and $d_3$
overlap. However, we note that $d_3 \subseteq d_4$ and then
$\omega(\{d_3,d_4\}) = \omega(\{d_4\})$ and $\{d_4\}$ is independent
from $\{d_1,d_2\}$.\punto
\end{example}

\subsection{Set Operations on ws-sets}
\label{sec:setop}
Various relevant computation tasks, ranging from decision procedures
like tuple possibility~\cite{AKG1991} to confidence computation of
answer tuples, and conditioning of probabilistic databases, require
symbolic manipulations of ws-sets. For example, checking whether two
tuples of a probabilistic relation can co-occur in some worlds can be
done by intersecting their ws-descriptors; both tuples co-occur in the
worlds defined by the intersection of the corresponding world-sets. 

We next define set operations on ws-sets.
\begin{itemize}
\item
{\it Intersection}. $\mbox{Intersect}(S_1, S_2) :=$

$\{d_1\cap
d_2\mid d_1\in S_1, d_2\in S_2, d_1 \mbox{ is consistent with } d_2 \}.$

\item
{\it Union}.
$\mbox{Union}(S_1, S_2) :=  S_1 \cup S_2$.

\item
{\it Difference}. The definition is inductive, starting with singleton ws-sets. 
If ws-descriptors $d_1$ and $d_2$ are inconsistent,
$
\mbox{Diff}(\{d_1\}, \{d_2\}) := \{d_1\}.
$
Otherwise,
\begin{multline*}
\mbox{Diff}(\{ d_1 \}, \{ d_2 \}) := \\
\{
d_1 \cup \{ x_1 \mapsto w_1, \dots, x_{i-1} \mapsto w_{i-1},
x_i \mapsto w'_{i} \}
\mid \\
d_2 - d_1 = 
\{ x_1 \mapsto w_1, \dots, x_k \mapsto w_k \}, \\
1 \le i \le k, w'_i \in \mbox{dom}_{x_i}, w_i \neq w'_i
\}.
\end{multline*}
\[
\mbox{Diff}(\{d_1\}, S \cup \{ d_2 \}) :=
\mbox{Diff}(\mbox{Diff}(\{d_1\}, S), \{ d_2 \}).
\]
\[
\mbox{Diff}(\{d_1, \dots, d_n\}, S) :=
\bigcup_{1 \le i \le n} \mbox{Diff}(\{d_i\}, S).
\]
\end{itemize}

\begin{example}\label{ex:diff}\em
Consider $d_1 = \{j\mapsto 1\}$, $d_2 = \{j\mapsto 7\}$, and
$d_3=\{j\mapsto 1, b\mapsto 4\}$.
Then, $\mbox{Intersect}(\{d_1\}, \{d_2\}) =
\mbox{Intersect}(\{d_2\}, \{d_3\}) = \emptyset$
because $d_2$ is inconsistent with $d_1$ and $d_3$.
$\mbox{Intersect}(\{d_1\}, \{d_3\}) = \{d_3\}$,
because $d_3$ is contained in $d_1$.
$\mbox{Diff}(\{d_2\}, \{d_1\}) = \mbox{Diff}(\{d_2\}, \{d_3\}) = \{d_2\}$
because $d_2$ is mutex with $d_1$ and $d_3$.
$\mbox{Diff}(\{d_1\}, \{d_3\}) = \{\{j\mapsto 1,b\mapsto 7\}\}$.
$\mbox{Diff}(\{d_3\}, \{d_1\}) = \{d_3\}$, because $d_3$ and $d_1$ are
inconsistent. \punto
\end{example}

\begin{proposition}\label{prop:diff-mutex}
The above definitions of set operations on ws-sets are correct:
\begin{enumerate}
\item
$\omega(\mbox{Union}(S_1, S_2)) = \omega(S_1) \cup \omega(S_2)$.
\item
$\omega(\mbox{Intersect}(S_1, S_2)) = \omega(S_1) \cap \omega(S_2)$.
\item
$\omega(\mbox{Diff}(S_1, S_2)) = \omega(S_1) - \omega(S_2)$.
\end{enumerate}

The ws-descriptors in $\mbox{Diff}(S_1, S_2)$ are pairwise mutex.
\end{proposition}

\begin{figure*}[t]
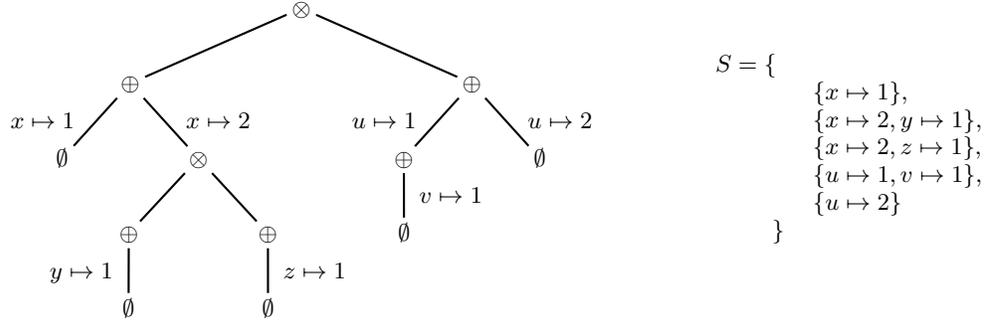


\begin{tabular}{l|ccc}
$W$ & V & D & P \\
\hline
& $x$ &  1 &   .1 \\
& $x$ &  2 &   .4 \\
& $x$ &  3 &   .5 \\
& $y$ &  1 &   .2 \\
& $y$ &  2 &   .8 \\
& $z$ &  1 &   .4 \\
& $z$ &  2 &   .6 \\
& $u$ &  1 &   .7 \\
& $u$ &  2 &   .3 \\
& $v$ &  1 &   .5 \\
& $v$ &  2 &   .5 \\
\end{tabular}\relax\parbox{0.6\textwidth}{\relax%
\[%
\psset{levelsep=10mm, nodesep=2pt, treesep=16mm}
\pstree{\TR{\otimes}}
{
  \pstree{\TR{\oplus}}
  {
    \TR{\emptyset}^{x \mapsto 1}
    \pstree{\TR{\otimes}_{x \mapsto 2}}
    {
      \pstree{\TR{\oplus}}
      {
        \TR{\emptyset}^{y \mapsto 1}
      }
      \pstree{\TR{\oplus}}
      {
        \TR{\emptyset}_{z \mapsto 1}
      }
    }
  }
  \pstree{\TR{\oplus}}
  {
    \pstree{\TR{\oplus}^{u \mapsto 1}}
    {
      \TR{\emptyset}_{v \mapsto 1}
    }
    \TR{\emptyset}_{u \mapsto 2}
  }
}
\]}\relax\parbox{0.4\textwidth}{\relax%
\begin{tabular}{ll}
$S = \{$ & \\
         & $\{ x \mapsto 1 \}$, \\
         & $\{ x \mapsto 2, y \mapsto 1 \}$, \\
         & $\{ x \mapsto 2, z \mapsto 1 \}$, \\
         & $\{ u \mapsto 1, v \mapsto 1 \}$, \\
         & $\{ u \mapsto 2 \}$ \\
\hspace*{2em} $\}$ & \\
\end{tabular}
}

\caption{World-set table $W$, a ws-tree $R$ over $W$, and an equivalent ws-set $S$.}
\label{fig:ws-tree}
\end{figure*}

\section{World-set trees}
\label{sec:ws-trees}

The ws-sets have important properties, like succinctness, closure
under set operations, and natural relational encoding, and \cite{AJKO07}
employed them to achieve the purely relational processing of
positive relational algebra on U-relational
da\-ta\-bases. When it comes to the manipulation of probabilities
of query answers or of worlds violating given constraints, however,
ws-sets are in most cases inadequate. This is because
ws-descriptors in a ws-set may represent non-disjoint world-sets, and
for most manipulations of probabilities a substantial computational
effort is needed to identify common world-subsets across
possibly many ws-descriptors.

We next introduce a new compact representation of world-sets, called
{\em world-set tree}\/ representation, or ws-tree for short, that makes the
structure in the ws-sets explicit. This representation formalism
allows for efficient exact probability computation and conditioning
and has strong connections to
\textit{knowledge compilation}, as it is used in system modelling and
verification~\cite{darwicheJAIR02}. There, too, various kinds of
decision diagrams, like binary decision diagrams
(BDDs)~\cite{Bryant86}, are employed for the efficient manipulation of
propositional formulas.

\begin{definition}\em
Given a world-table $W$, a ws-tree over $W$ is a tree with inner nodes
$\otimes$ and $\oplus$, leaves holding the ws-descriptor $\emptyset$,
and edges annotated with weighted variable assignments consistent with
$W$. The following constraints hold for a ws-tree:
\begin{itemize}
\item A variable defined in $W$ occurs at most once on each root-to-leaf path.
\item Each of its $\oplus$-nodes is associated with a variable $v$ such that 
each outgoing edge is annotated with a different assignment of $v$.
\item The sets of variables occurring in the subtrees rooted at the children
 of any $\otimes$-node are disjoint.\punto
\end{itemize}
\end{definition}

We define the semantics of ws-trees in strict analogy to that of
ws-sets based on the observation that the set of edge annotations on
each root-to-leaf path in a ws-tree represents a ws-descriptor. The
world-set represented by a ws-tree is precisely represented by the
ws-set consisting of the annotation sets of all root-to-leaf paths.
The inner nodes have a special semantics: the children of a
$\otimes$-node use disjoint variable sets and are thus independent,
and the children of a $\oplus$-node follow branches with different
assignments of the same variable and are thus mutually exclusive.

\begin{example}\em
Figure~\ref{fig:ws-tree} shows a ws-tree and the ws-set consisting of
all its root-to-leaf paths. \punto
\end{example}


\begin{figure}[t!]
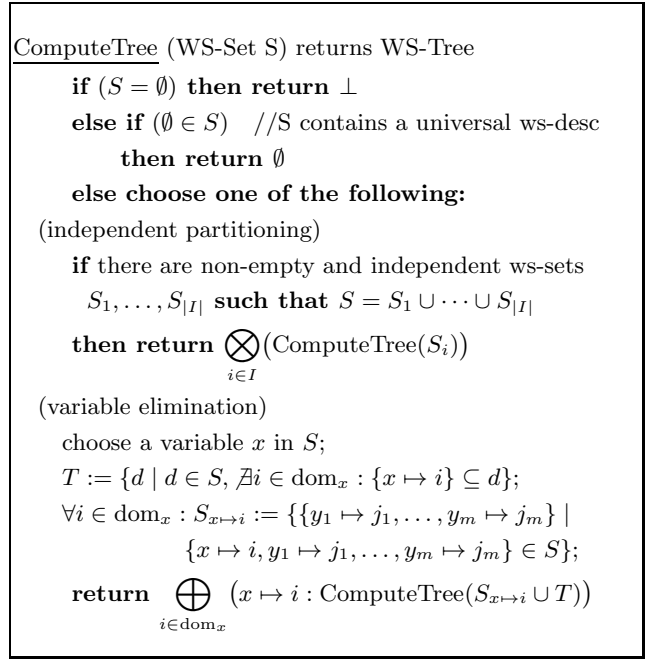

\framebox[\columnwidth]{
\parbox{8cm}{
\begin{align*}
&\hspace*{-2em} \underline{\mbox{ComputeTree}} \mbox{ (WS-Set S) returns WS-Tree}\\
&\mbox{\bf\ if } (S = \emptyset) \mbox{\bf\ then return } \bot\\
&\mbox{\bf\ else if } (\emptyset\in S) \hspace*{1em}\mbox{//S contains a universal ws-desc}\\
&\hspace*{2em} \mbox{\bf\ then return } \emptyset\\
&\mbox{\bf\ else choose one of the following:}\\
&\hspace*{-1em} \mbox{(independent partitioning)}\\
&\mbox{\bf\ if } \mbox{there are non-empty and independent ws-sets}\\
&\hspace*{1em}  S_1, \dots, S_{|I|} \mbox{\bf\ such that } S=S_1 \cup \dots \cup S_{|I|}\\
&\mbox{\bf\ then return } \underset{i\in I}{\bigotimes} \big( \mbox{ComputeTree}(S_i) \big)\\
&\hspace*{-1em} \mbox{(variable elimination)}\\
&\mbox{choose a variable } x \mbox{ in } S; \\
&T := \{d\mid d\in S, \not\exists i\in\mathrm{dom}_x: \{x\mapsto i\}\subseteq d\};\\
&\forall i\in \mathrm{dom}_x: S_{x\mapsto i} := \{\{y_1\mapsto j_1,\ldots,y_m\mapsto j_m\} \mid\\
&\hspace*{5em} \{x\mapsto i,y_1\mapsto j_1,\ldots,y_m\mapsto j_m\} \in S\};\\
&\mbox{\bf\ return } \underset{i\in\mathrm{dom}_x}{\bigoplus} \big( x\mapsto i: \mbox{ComputeTree}(S_{x\mapsto i}\cup T)\big)
\end{align*}
 \centering
}}
 \caption{Translating ws-sets into ws-trees.}
 \label{fig:translation}
\end{figure}

\subsection{Constructing world-set trees}
\label{sec:translation}

The key idea underlying our translation of ws-sets into ws-trees is a
divide-and-conquer approach that exploits the relationships between
ws-descriptors, like independence and variable sharing.

Figure~\ref{fig:translation} gives our translation algorithm. We
proceed recursively by partitioning the ws-sets into independent
disjoint partitions (when possible) or into (possibly overlapping)
partitions that are consistent with different assignments of a
variable. In the case of independent partitioning, we create
$\otimes$-nodes whose children are the translations of the independent
partitions. In the second case, we simplify the problem by eliminating
a variable: we choose a variable $x$ and create an $\oplus$-node whose
outgoing edges are annotated with different assignments $x\mapsto i$
of $x$ and whose children are translations of the subsets of the
ws-set consisting of ws-descriptors consistent with $x\mapsto
i$\footnote{Our translation abstracts out implementation details. For
instance, for those assignments of $x$ that do not occur in $S$ we
have $T\cup S_{x\mapsto i} = T$ and can translate $T$ only once.}. If
at any recursion step the input ws-set contains the nullary
ws-descriptor, which by definition represents the whole world-set,
then we stop from recursion and create a ws-tree leaf
$\emptyset$. This can happen after several variable elimination steps
that reduced some of the input ws-descriptors to $\emptyset$.

\begin{example}\em
We show how to translate the ws-set $S$ into the ws-tree $R$
(Figure~\ref{fig:ws-tree}). We first partition $S$ into two
(minimally) independent ws-sets $S_1$ and $S_2$: $S_1$ consists of the
first three ws-descriptors of $S$, and $S_2$ consists of the remaining
two. For $S_1$, we can eliminate any of the variables $x$, $y$, or
$z$. Consider we choose $x$ and create two branches for $x\mapsto 1$
and $x\mapsto 2$ respectively (there is no ws-descriptor consistent
with $x\mapsto 3$). For the first branch, we stop with the ws-set
$\{\emptyset\}$, whereas for the second branch we continue with the
ws-set $\{\{y\mapsto 1\},\{z\mapsto 1\}\}$. The latter ws-set can be
partitioned into independent subsets in the
\textit{context} of the assignment $x\mapsto 2$. We
proceed similarly for $S_2$ and choose to eliminate variable $u$.  We
create an $\oplus$-node with outgoing edges for assignments $u\mapsto
1$ and $u\mapsto 2$ respectively. We are left in the former case with
the ws-set $\{\{v\mapsto 1\}\}$ and in the latter case with
$\{\emptyset\}$.

Different variable choices can lead to different ws-trees. This is the
so-called \textit{variable ordering problem} that applies to the
construction of binary decision diagrams. Later in this section we
discuss heuristics for variable orderings. \punto
\end{example}

\begin{theorem}\label{th:ws-tree}
Given a ws-set $S$, \mbox{ComputeTree}($S$) and $S$ represent the same
world-set.
\end{theorem}

Our translation can yield ws-trees of exponential size (similar to
BDDs). This rather high worst-case complexity needs to be paid for
efficient exact probability computation and conditioning. It is known
that counting models of propositional formulas and exact
probability computation are \#P-hard problems~\cite{dalvi07efficient}.
This complexity
result does not preclude, however, BDDs from being very successful in
practice. We expect the same for ws-trees. The key observation for a
good behaviour in practice is that we should partition ws-sets into
independent subsets whenever possible and we should carefully choose a
good ordering for variable eliminations. Both methods greatly
influence the size of the ws-trees and the translation time, as shown
in the next example.

\begin{figure}

{\footnotesize
\[%
\psset{levelsep=8mm, nodesep=2pt, treesep=11mm}
  \pstree{\TR{\oplus}}
  {
   \pstree{\TR{\oplus}^{y \mapsto 1}}
    {
      \pstree{\TR{\otimes}^{u \mapsto 1}}
       {
         \pstree{\TR{\oplus}}
          {
            \TR{\emptyset}^{v \mapsto 1}
          }
        \pstree{\TR{\oplus}}
          {
            \pstree{\TR{\oplus (\alpha)}^{z \mapsto 2}}
             {
              \TR{\emptyset}^{x \mapsto 1}
              \TR{\emptyset}_{x \mapsto 2}
             }
            \TR{\alpha}
          }
       }
     \TR{\emptyset}_{u \mapsto 2}
    }
   \pstree{\TR{\oplus}_{y \mapsto 2}}
    {
      \TR{\emptyset}^{x \mapsto 1}
      \pstree{\TR{\otimes}_{\hspace*{-2em} x \mapsto 2}}
       {
         \pstree{\TR{\oplus}}
          {
            \TR{\emptyset}^{z \mapsto 1\mbox{  }}
          }
         \pstree{\TR{\oplus (\beta)}}
          {
          \TR{\emptyset}^{u \mapsto 2}
           \pstree{\TR{\oplus}_{u \mapsto 1}}
            {
              \TR{\emptyset}_{v \mapsto 1}
	    }
	  }
       }
      \TR{\beta}_{\mbox{   }x \mapsto 3}
    }
}
\]
}

\caption{A ws-tree equivalent to $R$ of Figure~\ref{fig:ws-tree}.}
\label{fig:ws-tree2}
\end{figure}

\begin{example}\em
Consider again the ws-set $S$ of Figure~\ref{fig:ws-tree} and a
different ordering for variable eliminations that leads to the ws-tree
of Figure~\ref{fig:ws-tree2}. We shortly discuss the construction of
this ws-tree. Assume we choose to eliminate the variable $y$ and
obtain the ws-sets
\begin{align*}
S_{y\mapsto 2} &= \{\{x\mapsto 1\}, \{x\mapsto 2,z\mapsto
1\},\{u\mapsto 1,v\mapsto 1\},\{u\mapsto 2\}\}\\
S_{y\mapsto 1} &= S_{y\mapsto 2}\cup\{\{x\mapsto 2\}\}
\end{align*}
In contrast to the computation of the ws-tree $R$ of
Figure~\ref{fig:ws-tree}, our variable choice creates intermediary
ws-sets that overlap at large, which ultimately leads to a large
increase in the size of the ws-tree. This bad choice need not
necessarily lead to \textit{redundant} computation, which we could
easily detect. In fact, the only major savings in case we detect and
eliminate redundancy here are the subtrees $\alpha$ and $\beta$, which
still leave a graph larger than $R$. \punto
\end{example}

\begin{figure}[t]
\framebox[\columnwidth]{
\hspace{2mm}
\parbox{8cm}{
\begin{tabular}{l}
\hspace*{-1.5em}\underline{Estimate} (WS-Set $S$, variable $x$ in $S$) returns Real \\[1ex]
missing\_assignment := false;\\
\textbf{foreach} $i\in\mathrm{dom}_x$ \textbf{do}\\
\hspace*{1em} compute $S_{x\mapsto i}$ and $T$ as shown in Figure~\ref{fig:translation}\\
\hspace*{1em} \textbf{if} $|S_{x\mapsto i}|>0$ \textbf{then} $s_i = |S_{x\mapsto i}\cup T|$\\
\hspace*{1em} \textbf{else} $s_i = 0$; missing\_assignment = true; \textbf{endif}\\
\textbf{if} (missing\_assignment) \textbf{then} $e = |T|$ \textbf{else} $e = 0$\\
\textbf{foreach} $j\in\mathrm{dom}_x$ \textbf{such that} $s_j>0$ \textbf{do}\\
\hspace*{1em} $e = e + \log_k(1 + k^{s_j-e})$\\
\textbf{return} $e$
\end{tabular}
 \centering
}
\hspace{2mm}
}
 \caption{Log cost estimate for a variable choice.}
 \label{fig:log-estimate}

\end{figure}

\subsection{Heuristics}
\label{sec:heuristics}

We next study heuristics for variable elimination and independent
partitioning that are compared experimentally in
Section~\ref{sec:experiments}. We devise a simple cost estimate, which
we use to decide at each step whether to partition or which variable
to eliminate. We assume that, in worst case, the cost of translating a
ws-set $S$ is $2^{|S|}$ (following the exponential formula of the
inclusion-exclusion principle).

In case of independent partitioning, the partitions $S_1,\ldots,$
$S_n$ are disjoint and can be computed in polynomial time (by
computing the connected components of the graph of variables
co-occurring within ws-descriptors). We thus reduce the computation
cost from $2^{|S|}$ to $2^{|S_1|}+\cdots+2^{|S_n|}$. This method is,
however, not always applicable and we need to apply variable
elimination.

The main advantage of variable elimination is that $S$ is divided into
subsets $T\cup S_{x\mapsto i}$ without the dependencies enforced by
variable $x$ and thus subject to independent partitioning in the
context of $x\mapsto i$. Consider $s_i$ the size of the ws-set $T\cup
S_{x\mapsto i}$. Then, the cost of choosing $x$ is
$\underset{i\in\mathrm{dom}_x}{\Sigma} 2^{s_i}$. Of course, for those
assignments of $x$ that do not occur in $S$ we have $T\cup S_{x\mapsto
i} = T$ and can translate $T$ only once. The computation cost using
variable elimination can match that of independent partitioning only
in the case that the assignments of the chosen variable partition the
input ws-set $S$ and thus $T$ is empty.

Our first heuristic, called \textit{minlog}, chooses a variable that
minimizes $\log(\underset{i=1}{\overset{\mathrm{dom}_x}{\Sigma}}
2^{s_i})$. Figure~\ref{fig:log-estimate} shows how to compute
incrementally the cost estimate by avoiding summation of potentially
large numbers. The variable missing\_assignment is used to detect
whether there is at least one assignment of $x$ not occuring in $S$
for which $T$ will be translated; in this case, $T$ is only translated
once (and not for every missing assignment).

The second heuristic, called \textit{minmax}, approximates the cost
estimate and chooses a variable that minimizes the maximal ws-set
$T\cup S_{x\mapsto i}$. Both heuristics need time linear in the sizes
of all variable domains plus of the ws-set. In addition to minmax,
minlog needs to perform log and exp operations.

\begin{remark}\em
To better understand our heuristics, we give one scenario where minmax
behaves suboptimal. Consider $S$ of size $n$ and two
variables. Variable $x$ occurs with the same assignment in $n-1$
ws-descriptors and thus its minmax estimate is $n$, and variable $y$
occurs twice with different assignments, and thus its minmax estimate
is $n-1$. Using minmax, we choose $y$, although the minlog would
choose differently: $e(y) = \log (2\cdot 2^{n-1} + 2^{n-2}) >
\log (2\cdot 2^{n-1}) = e(x)$.\punto
\end{remark}

\begin{figure}
\framebox[\columnwidth]{
\hspace{2mm}
\parbox{8cm}{
\begin{align*}
P \big(\bigotimes_{i \in I} S_i\big) & = 1 - \prod_{i \in I} (1 - P(S_i))\\
P \big( \bigoplus_{i \in I} (x \mapsto i : S_i) \big) &= \sum_{i \in I} P(\{x \mapsto i\}) \cdot P(S_i)\\
P (\emptyset) &= 1 \hspace*{4em} P (\bot) = 0
\end{align*}
}}

\caption{Probability computation for ws-trees.}
\label{fig:ws-tree-prob}
\end{figure}

\subsection{Probability computation}
\label{sec:prob-computation}

We next give an algorithm for computing the exact probability of a
ws-set by employing the translation of ws-sets into ws-trees discussed
in Section~\ref{sec:ws-trees}. Figure~\ref{fig:ws-tree-prob} defines
the function $P$ to this effect. This function is defined using
pattern matching on the node types of ws-trees. The probability of an
$\otimes$-node is the joint probability of its independent children
$S_1,\ldots,S_{|I|}$. The probability of an $\oplus$-node is the joint
probability of its mutually exclusive children, where the probability
of each child $S_i$ is weighted by the probability of the variable
assignment $x\mapsto i$ annotating the incoming edge of
$S_i$. Finally, the probability of a leaf represented by the nullary
ws-descriptor is 1 and of $\bot$ is 0.

\begin{example}\em\label{ex:prob}
The probability of the ws-tree $R$ of Figure~\ref{fig:ws-tree} can be
computed as follows (we label the inner nodes with $l$ for left child
and $r$ for right child):
\begin{align*}
P(R) &= 1 - (1 - P(l))\cdot(1 - P(r))\\
P(l) &= P(\{x\mapsto 1\})\cdot P(\emptyset) + P(\{x\mapsto 2\})\cdot P(lr)\\
P(lr) &= 1 - (1 - P(\{y\mapsto 1\})\cdot P(\emptyset))\cdot (1 - P(\{z\mapsto 1\})\cdot P(\emptyset))\\
P(r) &= P(\{u\mapsto 1\})\cdot P(\{v\mapsto 1\})\cdot P(\emptyset) + P(\{u\mapsto 2\})\cdot P(\emptyset)
\end{align*}
We can now replace the probabilities for variable assignments and
ws-descriptor $\emptyset$ and obtain
\begin{align*}
P(r) &= 0.7\cdot 0.5\cdot 1 + 0.3 = 0.65\\
P(lr) &= 1 - (1 - 0.2\cdot 1)\cdot (1 - 0.4\cdot 1) = 0.52\\
P(l) &= 0.1\cdot 1 + 0.4\cdot 0.52 = 0.308\\
P(R) &= 1 - (1 - 0.308)\cdot (1 - 0.65) = 0.7578 \hspace*{6em}\Box
\end{align*}
\end{example}

The probability of a ws-tree $R$ can be computed in one bottom-up
traversal of $R$ and does not require the precomputation of $R$. The
translation and probability computation functions can be easily
composed to obtain the function $\mbox{ComputeTree}\circ P$ by
inlining $P$ in $\mbox{ComputeTree}$. As a result, the construction of
the nodes $\oplus$, $\otimes$, and $\emptyset$ is replaced by the
corresponding probability computation given by $P$.

\section{Conditioning}
\label{sect:excltab}

In this section we study the problem of conditioning a probabilistic
database, i.e., the problem of removing all possible worlds that
do not satisfy a given condition (say, by a Boolean relational calculus
query) and renormalizing the database such that, if there is at least
one world left, the probability weights of all worlds sum up to one.

We will think of conditioning as a query or update
operation assert$_\phi$, where
$\phi$ is the condition, i.e., a Boolean query.
Processing relational algebra queries on
probabilistic databases was discussed in Section~\ref{sec:pdb}.
We will now assume the result of the Boolean query given as a
ws-set defining the worlds on which $\phi$ is true.

\begin{example}\label{ex:conditioning1}\em
Consider again the data cleaning example from the Introduction, formalized
by the U-relational database of Figure~\ref{fig:pdb}.
Relation $W$ represents the set of possible worlds and $U$ represents the
tuples in these worlds.

As discussed in Example~\ref{ex:fd}, the set of ws-descriptors
$S = \{ \{ j \mapsto 1 \}, \{ j \mapsto 7, b \mapsto 4 \} \}$
represents the three worlds on which the functional dependency
SSN $\rightarrow$ NAME holds.
The world $\{ j \mapsto 7, b \mapsto 7 \}$ is excluded and thus the confidence
of $S$ does not add up to one but to $.2 + .8 \cdot .3 = .44$.
What we now want to do is
transform this database into one that represents the three worlds identified
by $S$ and preserves their tuples as well as their relative weights, but
with a sum of world weights of one. This can of course be easily achieved
by multiplying the weight of each of the three remaining worlds by
1/.44. However, we want to do this in a smart way that in general does
not require to consider each possible world individually, but instead
preserves a succinct representation of the data and runs efficiently.

Such a technique exists and is presented in this section. It is based
on running our confidence computation algorithm for ws-trees and,
while returning from the recursion, renormalizing the world-set by
introducing new variables whose assignments are normalized using the
confidence values obtained.  For this example, the conditioned
database will be
\[
\begin{tabular}{l|ccc}
$W$ & Var & Dom & P \\
\hline
    & $b$ & 4  & .3   \\
    & $b$ & 7  & .7   \\
    & $j'$ & 1 & .2/.44 \\
    & $j'$ & 7 & $.8 \cdot .3/.44$ \\
\end{tabular}
\]
\[
\begin{tabular}{l|l|ll}
U & WSD & SSN & NAME \\
\hline
  & $\{ j' \mapsto 1 \}$ & 1 & John \\
  & $\{ j' \mapsto 7 \}$ & 7 & John \\
  & $\{ j' \mapsto 1, b \mapsto 4 \}$ & 4 & Bill \\
  & $\{ j' \mapsto 1, b \mapsto 7 \}$ & 7 & Bill \\
  & $\{ j' \mapsto 7 \}$ & 4 & Bill \\
\end{tabular}
\]
Note that the $W$ relation actually models four possible worlds, but
two of them, $\{ j' \mapsto 7, b \mapsto 4 \}$ and $\{ j' \mapsto 7, b
\mapsto 7 \}$ are equal (contain the same tuples). Example~\ref{ex:conditioning2} 
will show in detail how conditioning works. \punto
\end{example}

\begin{figure}[t]
\framebox[\columnwidth]{
\hspace*{-2em}
\parbox{8cm}{
\begin{tabular}{l@{\hspace*{0.2em}}l}
\underline{cond}: & conditioning algorithm\\ 
In: & ws-tree $R$ representing the new nonempty world-set, \\ 
    & ws-set $U$ from the U-relations \\ 
Out: & (confidence value, ws-set $U'$) \\[1ex]
%
%
& \textbf{if} $R = \emptyset$ \textbf{then} \textbf{return} $(1,U)$;\\[1ex]
& \textbf{if} $R = \underset{i\in I}{\bigotimes} \big( R_i \big)$ \textbf{then}\\
& \hspace*{1em} \textbf{foreach} $i\in I$ \textbf{do} $(c_i,U_i) := \mbox{cond}(R_i,U)$;\\
& \hspace*{1em} \textbf{return} $(1 - \prod_i(1 - c_i),\underset{i\in I}{\bigcup} U_i)$;\\[1ex]
& \textbf{if} $R = \underset{i\in \mbox{dom}_x}{\bigoplus} \big( x\mapsto i: R_i \big)$ \textbf{then}\\
& \hspace*{1em} \textbf{foreach} $i\in \mbox{dom}_x$ \textbf{do} \\
& \hspace*{2em} $U_i :=$  the subset of $U$ consistent with $x\mapsto i$;\\
& \hspace*{2em} $(c_i,U'_i) := \mbox{cond}(R_i,U_i)$;\\
& \hspace*{1em} $c := \sum_{i \in I} P(\{x \mapsto i\}) \cdot c_i$;\\
& \hspace*{1em} let $x'$ be a new variable; \\
& \hspace*{1em} \textbf{foreach} $i\in \mbox{dom}_x$ \textbf{such that} $c_i\not=0$ \textbf{do} \\
& \hspace*{2em} add $\tuple{x', i, \frac{P(\{x \mapsto i\}) \cdot c_i}{c}}$ to the $W$ relation; \\
& \hspace*{2em} replace each occurrence of $x$ in $U_i'$ by $x'$; \\
& \hspace*{1em} {\bf return} $(c,\underset{i\in \mbox{dom}_x}\bigcup U_i')$;
\end{tabular}
}}

\caption{The conditioning algorithm.}
\label{fig:conditioning_algorithm}
\end{figure}

\psset{levelsep=10mm, nodesep=2pt, treesep=16mm}

\begin{figure*}[t]
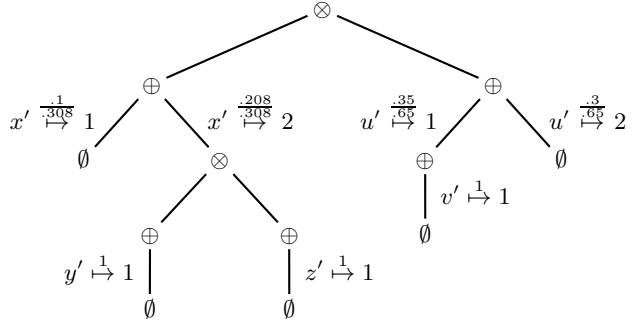

\parbox{0.4\textwidth}{
\[
\begin{tabular}{c}
\begin{tabular}{l|l|l}
$U$ & WSD & A \\
\hline
    & $\{ y \mapsto 2, u \mapsto 1 \}$ & $a_1$ \\
    & $\{ u \mapsto 1, v \mapsto 2 \}$ & $a_2$ \\
\end{tabular}\\[2em]
\begin{tabular}{l|ccc}
$\Delta W$ & Var & Dom & P \\
\hline
& $x'$ &  1 &   .1/.308 \\
& $x'$ &  2 &   .208/.308 \\
& $y'$ &  1 &   1 \\
& $z'$ &  1 &   1 \\
& $u'$ &  1 &   .35/.65 \\
& $u'$ &  2 &   .3/.65 \\
& $v'$ &  1 &   1 \\
\end{tabular}%
\end{tabular}\]}\relax\parbox{0.6\textwidth}{\relax%
\[%
\pstree{\TR{\otimes}}
{
  \pstree{\TR{\oplus}}
  {
    \TR{\emptyset}^{x' \stackrel{\frac{.1}{.308}}{\mapsto} 1}
    \pstree{\TR{\otimes}_{x' \stackrel{\frac{.208}{.308}}{\mapsto} 2}}
    {
      \pstree{\TR{\oplus}}
      {
        \TR{\emptyset}^{y' \stackrel{1}{\mapsto} 1}
      }
      \pstree{\TR{\oplus}}
      {
        \TR{\emptyset}_{z' \stackrel{1}{\mapsto} 1}
      }
    }
  }
  \pstree{\TR{\oplus}}
  {
    \pstree{\TR{\oplus}^{u' \stackrel{\frac{.35}{.65}}{\mapsto} 1}}
    {
      \TR{\emptyset}_{v' \stackrel{1}{\mapsto} 1}
    }
    \TR{\emptyset}_{u' \stackrel{\frac{.3}{.65}}{\mapsto} 2}
  }
}
\]
}
\vspace{-4mm}

\caption{U-relation $U$, additions $\Delta W$ to the $W$-relation, and a renormalized ws-tree.}
\label{fig:conditioning_example}
\end{figure*}

Figure~\ref{fig:conditioning_algorithm} gives our efficient algorithm
for conditioning a U-rela\-tion\-al database. The input is a
U-relational database and a ws-tree $R$ that describes the subset of
the possible worlds of the database that we want to condition it
to. The output is a modified U-relational database and, as a
by-product, since we recursively need to compute confidences for the
renormalization, the confidence of $R$ in the input database. The
confidence of $R$ in the output database will of course be 1. The
renormalization works as follows. The probability of each branch of an
inner node $n$ of $R$ is re-weighted such that the probability of $n$
becomes 1. We reflect this re-weighting by introducing new variable
whose assignments reflect the new weights of the branches of $n$.

This algorithm is essentially the confidence computation algorithm of
Figure~\ref{fig:ws-tree-prob}. We just add some lines of code along
the line of recursively computing confidence that renormalize the
weights of alternative assignments of variables for which some
assignments become impossible. Additionally, we pass around a set of
ws-descriptors (associated with tuples from the input U-relational
database) and extend each ws-descriptor in that set by $x \mapsto i$
whenever we eliminate variable $x$, for each of its alternatives $i$.

\begin{example}\label{ex:conditioning2}\em
Consider the U-relational database consisting of the $W$-relation of
Figure~\ref{fig:ws-tree} and the U-relation $U$ of
Figure~\ref{fig:conditioning_example}.  Let us run the algorithm to
condition the database on the ws-tree $R$ of Figure~\ref{fig:ws-tree}
($R$ need not be precomputed for conditioning).

We recursively call function cond at each node in the ws-tree $R$
starting at the root. To simplify the explanation, let us assume a
numbering of the nodes and of the ws-sets we pass around: If $R_w$ is
a (sub)tree then $R_{w,i}$ is its $i$-th child. The ws-set passed in
the recursion with $R_w$ is $U_w$ and the ws-set returned is $U'_w$.
The ws-sets passed on at the nodes of $R$ are:
\begin{align*}
U_1 & = U_2 = U\\
U_{1,1}   & = x \mapsto 1 : U = \{ \{ x \mapsto 1, y \mapsto 2, u \mapsto 1 \},\\ 
          & \hspace*{2em}\{ x \mapsto 1, u \mapsto 1, v \mapsto 2\} \}\\
U_{1,2}   &= x \mapsto 2 : U = \{ \{ x \mapsto 2, y \mapsto 2, u \mapsto 1 \},\\
          & \hspace*{2em}\{ x \mapsto 2, u \mapsto 1, v \mapsto 2 \} \}\\
U_{1,2,1,1} &= y\mapsto 1: U_{1,2} = \{\{ y\mapsto 1, x \mapsto 2, u \mapsto 1, v \mapsto 2 \}\}\\
U_{1,2,2,1} &= z\mapsto 1: U_{1,2} = \{\{ z\mapsto 1, x \mapsto 2, y \mapsto 2, u \mapsto 1 \},\\
            & \hspace*{2em}\{ z\mapsto 1, x \mapsto 2, u \mapsto 1, v \mapsto 2 \}\}\\
U_{2,1} &= u\mapsto 1: U_2 = U_2\\
U_{2,2} &= u\mapsto 2: U_2 = \emptyset\\
U_{2,1,1} &= v\mapsto 1: U_{2,1} = \{\{ v\mapsto 1, y \mapsto 2, u \mapsto 1 \}\}
\end{align*}

When we reach the leaves of $R$, we start returning from recursion and
do the following. We first compute the probabilities of the nodes of
$R$ -- in this case, they are already computed in
Example~\ref{ex:prob}. Next, for each $\oplus$-node representing the
elimination of a variable, say $\alpha$, we create a new variable
$\alpha'$ with the assignments of $\alpha$ present at that node. In
contrast to $\alpha$, the assignments of $\alpha'$ are re-weighted by
the probability of that $\oplus$-node so that the sum of their weights
is 1. The new variables and their weighted assignments are given in
Figure~\ref{fig:conditioning_example} along the original ws-tree $R$
and in the $\Delta W$ relation to be added to the world table $W$.

When we return from recursion, we also compute the new ws-sets $U'_i$
from $U_i$. These ws-sets are equal in case of leaves and
$\otimes$-nodes, but, in case of $\oplus$-nodes, the variable
eliminated at that node is replaced by the new one we created.  In
case of $\oplus$ and $\otimes$ nodes, we also return the union of all
$U'_i$ of their children. We finally return from the first call with
the following ws-set $U'$:
\begin{align*}
\{ &\{ x' \mapsto 1, y \mapsto 2, u \mapsto 1 \},\\
   &\{ x' \mapsto 1, u \mapsto 1, v \mapsto 2 \},\\
   &\{ x' \mapsto 2, y' \mapsto 1, u \mapsto 1, v \mapsto 2 \},\\
   &\{ x' \mapsto 2, z' \mapsto 1, y \mapsto 2, u \mapsto 1 \},\\
   &\{ x' \mapsto 2, z' \mapsto 1, u \mapsto 1, v \mapsto 2 \},\\
   &\{ u' \mapsto 1, v' \mapsto 1, y \mapsto 2 \} \}.
\end{align*}
\punto
\end{example}

Let us view a probabilistic database semantically, as a set of pairs
$(I,p)$ of instances $I$ with probability weights $p$.

\begin{theorem}\label{th:conditioning}
Given a representation of probabilistic da\-ta\-base ${\bf W} = \{
(I_1, p_1), \dots, (I_n, p_n) \}$ and a ws-tree $R$ identifying a
nonempty subset of the worlds of {\bf W}, the algorithm of
Figure~\ref{fig:conditioning_algorithm} computes a representation of
probabilistic database
\[
\{ (I_j, \frac{p_j}{c}) \mid (I_j, p_j) \in {\bf W}, I_j \in \omega(R) \} 
\]
such that the probabilities $p_j$ add up to 1.
\end{theorem}

Thus, of course, $c$ is the confidence of $R$.

Three simple optimizations of this algorithm that simplify the world
table $W$ and the output ws-descriptors are worth mentioning.
\begin{enumerate}
\item
Variables that do not appear anywhere in the U-relations can be
dropped from $W$.

\item
Variables with a single domain value (obviously of weight 1) can be
dropped everywhere from the database.

\item
Variables $x'$ and $x''$ obtained from the same variable $x$ (by
creation of a new variable in the case of variable elimination on $x$
in two distinct branches of the recursion) can be merged into the same
variable if the alternatives and their weights in the $W$ relation are
the same. In that case we can replace $x''$ by $x'$ everywhere in the
database.
\end{enumerate}

\begin{example}\em
In the previous example, we can remove the variables $y', z'$, and $v'$ from
the $W$-relation and all variable assignments involving these variables from
the U-relation because of (1).
Furthermore, we can remove the variables $x$ and $z$ because of (1).
The resulting database is
\[
\begin{tabular}{l|l|l}
$U'$ & WSD & A \\
\hline
& $\{ x' \mapsto 1, y \mapsto 2, u \mapsto 1 \}$ & $a_1$ \\
& $\{ x' \mapsto 1, u \mapsto 1, v \mapsto 2 \}$ & $a_2$ \\
& $\{ x' \mapsto 2, u \mapsto 1, v \mapsto 2 \}$ & $a_2$ \\
& $\{ x' \mapsto 2, y \mapsto 2, u \mapsto 1 \}$ & $a_1$ \\
& $\{ x' \mapsto 2, u \mapsto 1, v \mapsto 2 \}$ & $a_2$ \\
& $\{ u' \mapsto 1, y \mapsto 2 \}$           & $a_1$ \\
\end{tabular}
\]\[
\begin{tabular}{l|ccc}
$W'$ & Var & Dom & P \\
\hline
& $x'$ &  1 &   .1/.308 \\
& $x'$ &  2 &   .208/.308 \\
& $y$ &  1 &   .2 \\
& $y$ &  2 &   .8 \\
& $u$ &  1 &   .7 \\
& $u$ &  2 &   .3 \\
& $u'$ &  1 &   .35/.65 \\
& $u'$ &  2 &   .3/.65 \\
& $v$ &  1 &   .5 \\
& $v$ &  2 &   .5 \\
\end{tabular}
\]
\end{example}

Finally, we state an important property of conditioning (expressed by
the assert operation) useful for query optimization.

\begin{theorem}
Assert-operations commute with other asserts and the operations of
positive relational algebra.
\end{theorem}


\section{ws-descriptor elimination}

We next present an alternative to exact probability computation using
ws-trees based on the difference operation on ws-sets, called here
ws-descriptor elimination. The idea is to incrementally eliminate
ws-descriptors from the input ws-set. Given a ws-set $S$ and a
ws-descriptor $d_1$ in $S$, we compute two ws-sets: the original
ws-set $S$ without $d_1$, and the ws-set representing the difference
of $\{d_1\}$ and the first ws-set. The probability of $S$ is then the
sum of the probabilities of the two computed ws-sets, because the two
ws-sets are mutex, as stated below by function $P_w$:
\begin{align*}
P_w (\emptyset) &= 0 \hspace*{4em} P_w (\{\emptyset\}) = 1\\
P_w (S) &= P_w (\{d_2,\ldots,d_n\}) + \underset{d\in(\{d_1\}-\{d_2,\ldots,d_n\})}{\sum} P(d)
\end{align*}
The function $P$ computes here the probability of a ws-descriptor.

\begin{example}\em
Consider the ws-set $\{d_1,d_2,d_3\}$ of
Example~\ref{ex:properties}. The ws-descriptor $d_2$ is mutex with
both $d_1$ and $d_3$ and we can eliminate it: $P_w(\{d_1,d_2,d_3\}) =
P_w(\{d_1,d_3\})+P(d_2)$. We now choose any to eliminate $d_3$ and
obtain $P_w(\{d_1,d_3\}) = P_w(\{d_3\}-\{d_1\})+P(d_1) = P(d_1)$, as
explained in Example~\ref{ex:diff}. Thus $P_w(\{d_1,d_2,d_3\}) =
P(d_2) + P(d_1) = 1$.\punto
\end{example}

This method exploits the fact that the difference operation preserves
the mutex property and is world-set monotone.

\begin{lemma}\label{prop:difference}
The following equations hold for any ws-sets $S_1$, $S_2$, and $S_3$: 
\begin{eqnarray*}
\omega(S_1 - S_2)                      &\subseteq& \omega(S_1)\\
\omega(S_1) \cup \omega(S_2)           &=& \omega(S_1 - S_2) \cup \omega(S_2)\\
\emptyset                                      &=& \omega(S_1 - S_2) \cap \omega(S_2)\\
\omega(S_1)\cap\omega(S_2)=\emptyset   &\Rightarrow&  \omega(S_1-S_3)\cap\omega(S_2-S_3)=\emptyset
\end{eqnarray*}
\end{lemma}

The correctness of probability computation by ws-descriptor
elimination follows immediately from Lemma~\ref{prop:difference}.

\begin{theorem}\label{th:wsd-elim}
Given a ws-set $S$, the function $P_w$ computes the probability of
$S$.
\end{theorem}

As a corollary, we have that

\begin{corollary}[Theorem~\ref{th:wsd-elim}]\label{cor:mutex-set}
Any ws-set $\overset{n}{\underset{i=1}{\bigcup}}\{d_i\}$ has the
equivalent mutex ws-set $\overset{n-1}{\underset{i=1}{\bigcup}} (\{d_i\} -
\overset{n}{\underset{j=i+1}{\bigcup}} \{d_j\}) \cup \{d_n\}$.
\end{corollary}

Like the translation of ws-sets into ws-trees, this method can take
exponential time in the size of the input ws-set. Moreover, the
equivalent mutex ws-set given above can be exponential. On the
positive side, computing the exact probability of such mutex ws-sets
can be done in linear time. Additionally, the probability of
$\{d\}-S_d$ can be computed on the fly without requiring to first
generate all ws-descriptors in the difference ws-set. This follows
from the fact that the difference operation on ws-descriptors only
generates mutex and distinct ws-descriptors. After generating a
ws-descriptor from the difference ws-set we can thus add its
probability to a running sum and discard it before generating the next
ws-descriptor. The next section reports on experiments with an
implementation of this method.

\section{Experiments}
\label{sec:experiments}

\begin{figure*}
        \begin{center}
\begin{tabular}{|l|c||r|r|r||r|}
\hline
Query & Size of  & TPC-H       & \#Input      & Size of & User\hspace*{1em}         \\
      & ws-desc. & Scale\hspace*{.5em}       &  Vars\hspace*{.5em} & ws-set  & Time(s)      \\\hline
{\footnotesize 
$Q_1$: \textbf{select} true \textbf{from}
        customer c, orders o, lineitem l}
      &          & 0.01        & 77215        & 9836    & 5.10        \\
{\footnotesize \textbf{where} c.mktsegment $=$ 'BUILDING' \textbf{and} c.custkey $=$ o.custkey}
      & 3        & 0.05        & 382314       & 43498   & 99.76       \\
{\footnotesize  and o.orderkey $=$ l.orderkey \textbf{and} o.orderdate $>$ '1995-03-15'}
      &          & 0.10        & 765572       & 63886   &  356.56      \\\hline
{\footnotesize $Q_2$: \textbf{select} true \textbf{from} lineitem}
      &          & 0.01        & 60175        & 3029    & 0.20        \\
{\footnotesize \textbf{where} shipdate \textbf{between} '1994-01-01' \textbf{and} '1996-01-01'}
      & 1        & 0.05        & 299814       & 15545   & 8.24        \\
{\footnotesize \textbf{and} discount \textbf{between} '0.05' \textbf{and} '0.08' \textbf{and} quantity $<$ 24}
      &          & 0.10        & 600572       & 30948   &  33.68       \\\hline
\end{tabular}

    \end{center}
  \vspace*{-3mm}
  \caption{TPC-H scenario: Queries, data characteristics, and performance of INDVE(minlog).}
  \label{fig:tpch}
\end{figure*}

\begin{figure*}[t!]
  \begin{center}
    \vspace{-1em}
        \begin{tabular}{c@{\ }cc}
        \includegraphics[scale=1]{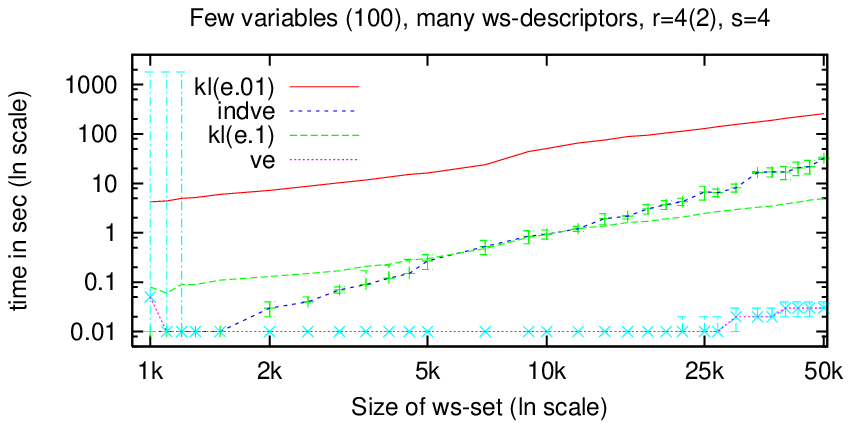}
        &
        \includegraphics[scale=1]{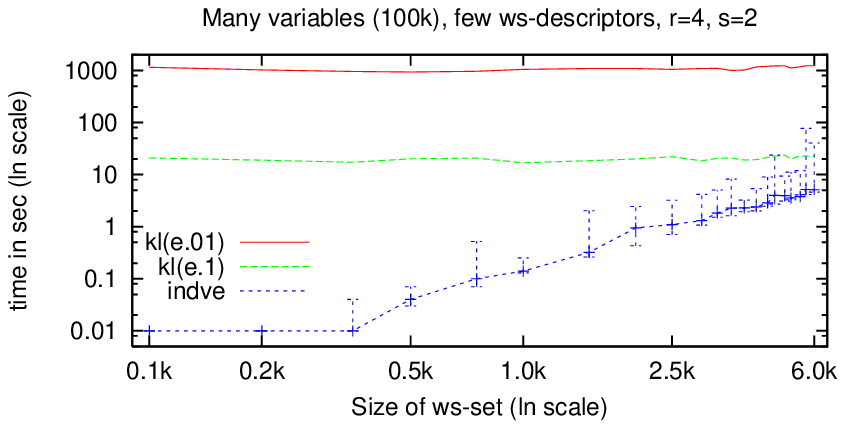}\\
        (a) & (b)
        \end{tabular}
  \end{center}
 \vspace*{-6mm}
 \caption{The two cases when the numbers of variables and of ws-descriptors differ by orders of magnitude.}
 \label{fig:diff}
 \vspace*{-1.5em}
\end{figure*}


The experiments were conducted on an Athlon-X2 (4600+)
x86-64bit/1.8GB/ Linux 2.6.20/gcc 4.1.2 machine.

\nop{
The code is available at
\begin{center}
\texttt{http://www.cs.cornell.edu/database/maybms}.
\end{center}
} 

\medskip

We considered two synthetic data sets.

\medskip

{\noindent\bf\underline{TPC-H data and queries}}. The first data set
consists of tuple-independent probabilistic databases obtained from
relational databases produced by
TPC-H 2.7.0, where each tuple is associated with a Boolean random
variable and the probability distribution is chosen at random.  We
evaluated the two Boolean queries of Figure~\ref{fig:tpch} on each
probabilistic database and then computed the probability of the ws-set
consisting of the ws-descriptors of all the answer tuples. Among the
two queries, only the second is \textit{safe} and thus admits PTIME
evaluation on tuple-independent probabilistic
databases~\cite{dalvi07efficient}. As we rewrite constraints into
Boolean queries, we consider this querying scenario equally relevant
to conditioning.

{\noindent\bf\underline{\#P-hard cases}}. The second data set consists
of ws-sets similar to those associated with the answers of
nonhierarchical conjunctive queries without self-joins on
tuple-independent probabilistic databases, i.e.\ join queries such as
$Q_s = R_1 \Join \dots \Join R_s$ for schemas $R_i(A_i, A_{i+1})$ in
which all relations are joined together, but there is no single column
common to all of them.  Such queries are known to be \#P-hard
\cite{dalvi07efficient}.

The data generation is simple: we partition the set of variables
into $s$ equally-sized sets $V_1, \dots, V_s$ and then sample ws-sets
$\{ x_1 \mapsto a_1, \dots, x_s \mapsto a_s \}$ where
$x_i$ is from $V_i$ and $a_i$ is a random alternative for $x_i$, for
$1 \le i \le s$. It is easy to verify that each such ws-set is actually
the result of query $Q_s$ on some tuple-independent probabilistic database.
(For $s=3$ this fact is used in the \#P-hardness proof of
\cite{dalvi07efficient}.)

We use the following parameters in our experiments: number $n$ of
variables ranging from 50 to 100K, number $r$ of possible alternatives
per variable (2 or 4), length $s$ of ws-descriptors, which equals the
number of joined relations (2 or 4), and number $w$ of ws-descriptors
ranging from 5 to 60K. For each variable, the alternatives have
uniform probabilities $1/r$: our exact algorithms are not sensitive to
changing probability values as long as the numbers of alternatives of
the variables remain constant.

\nop{
\begin{algorithm} (Data Generator) \em \\
Input: size $s$ of ws-descriptors,
       set $V$ of random variables of domain size $r$ ($|V|$ is a multiple
       of $s$), and
       size $w$ of output ws-set
\\
Output: ws-set $S$ of size $w$

\begin{enumerate}
\item
Partition $V$ into $s$ equally-sized sets $V_1,\ldots,V_s$.

\item
For each $V_i$
($1\leq i\leq s$), create a relation $R_i$ over schema $\Sigma_i$ such
that each tuple is associated with a random assignment of a variable of $V_i$.

\item
For each pair of relations $(R_i,R_{i+1})$ ($1\leq i\leq s-1$),
create a relation $P_{i,i+1}$ over schema $\Sigma_i\cup\Sigma_{i+1}$
that randomly pairs tuples from $R_i$ and $R_{i+1}$.

FROM CK: how many such tuples? We don't really do this, but first
creates sets $A_1, \dots, A_s$ of variable assignmens and then
take $w$ random tuples from $A_1 \times \dots \times A_s$, right?

\item
Perform $T := R_1\Join P_{1,2}\ldots\Join P_{s-1,s}\Join R_s$.

\item
Let ws-set $S$ be the set of the ws-descriptors of the first $w$ tuples of $T$.
\end{enumerate}
\end{algorithm}
} 

Note that the focus on Boolean queries means no loss of generality for
confidence computation; rather, the projection of a query result to
a nullary relation causes all the ws-sets to be unioned and large.

\medskip

{\noindent\bf\underline{Algorithms}}.
We experimentally compared three versions of our exact algorithm:
one that employs
independent partitioning and variable elimination (INDVE), one that
employs variable elimination only (VE), and one with ws-descriptor
elimination (WE). We considered INDVE with the two heu\-ristics minlog
and minmax.  These implementations compute confidence
values and the modified world table ($\Delta W$ in
Example~\ref{ex:conditioning2}),
but do not materialize the modified, conditioned U-relations
($U'$ in Example~\ref{ex:conditioning2}).
We have verified that the computation of these additional data
structures adds only a small overhead over confidence computation
in practice.  We therefore do not distinguish in the sequel
between confidence computation and conditioning.
Note that our implementation is based on the straightforward composition
of the ComputeTree and conditioning algorithms and does not need to
materialize the ws-trees.

Although we also implemented a brute-force algorithm for probability
computation, its timing is extremely bad and not reported. At a
glance, this algorithm iterates over all worlds and sums up the
probabilities of those that are represented by some ws-descriptors in
the input ws-set. We also tried a slight improvement of the
brute-force algorithm by first partitioning the input ws-set into
independent subsets~\cite{STW2008}. This version, too, performed bad
and is not reported, as the partitioning can only be applied once at
the beginning on the whole ws-set, yet most of our input ws-sets only
exhibit independence in the context of variable eliminations.

We experimentally compared INDVE against a Monte Carlo simulation algorithm for
confidence computation~\cite{RDS07,dalvi07efficient} which is based on
the Karp-Luby (KL) fully polynomial randomized approximation scheme (FPRAS) for DNF counting
~\cite{KL1983}. Given a DNF formula with $m$ clauses,
the base algorithm computes an $(\epsilon, \delta)$-approximation $\hat{c}$ of
the number of solutions $c$ of the DNF formula such that
\[
\Pr[|c - \hat{c}| \le \epsilon \cdot c] \ge 1 - \delta
\]
for any given $0 < \epsilon < 1$, $0 < \delta < 1$. It does so within
$\lceil 4 \cdot m \cdot \log(2/\delta) / \epsilon^2 \rceil$
iterations of an efficiently computable estimator.
This algorithm can be easily turned into an $(\epsilon, \delta)$-FPRAS
for tuple confidence computation (see \cite{Koch2008}).
In our experiments, we use the
optimal Monte-Carlo estimation algorithm of \cite{DKLR2000}.
This is a technique to determine a small sufficient number of Monte-Carlo
iterations (within a constant factor from optimal) based on first collecting
statistics on the input by running the Monte Carlo simulation a small number
of times.  We use the version of the Karp-Luby unbiased estimator described
in the book \cite{Vazirani2001}, which converges faster than the basic
algorithm of \cite{KL1983}, adapted to the problem of computing confidence
values. This algorithm is similar to the self-adjusting coverage
algorithm of \cite{KLM1989}.

\medskip

{\noindent\bf 1. Queries on TPC-H data.} 
Figure~\ref{fig:tpch} shows that INDVE(minlog) performs within
hundreds of seconds in case of queries with equi-joins ($Q_1$) and
selection-projection ($Q_2$) on tuple-independent probabilistic TPC-H
databases with over 700K variables and 60K ws-descriptors.
In the answers of query $Q_2$, ws-descrip\-tors are
pairwise independent, and
INDVE can effectively employ independence partitition, making confidence
computation more efficient than for $Q_1$.

\medskip

The remaining experiments use the second data generator.

\medskip

{\noindent\bf 2. The numbers of variables and of ws-descriptors differ
by orders of magnitude.} If there are much more ws-descriptors than
variables, many ws-descriptors share variables (or variable
assignments) and a good choice for variable elimination can
effectively partition the ws-set. On the other hand, independence
partitioning is unlikely to be very effective, and the time for
checking it is wasted. Figure~\ref{fig:diff}(a) shows that in such
cases VE and INDVE (with minlog heuristic) are very stable and not
influenced by fluctuations in data correlations. In particular, VE
performs better than INDVE and within a second for 100 variables with
domain size 4 (and nearly the same for 2), ws-descriptors of length 4,
and ws-set size above 1.2k. We witnessed a sharp hard-easy transition
at 1.2k, which suggests that the computation becomes harder when the
number of ws-descriptors falls under one order of magnitude greater
than the number of variables. Experiment 3 studies easy-hard-easy
transitions in more detail. The plot data were produced from 25 runs
and record the median value and ymin/ymax for the error bars.

In case of many variables and few ws-descriptors, the independence
partitioning clearly pays off. This case naturally occurs for query
evaluation on probabilistic databases, where a small set of tuples
(and thus of ws-descriptors) is selected from a large database. As
shown in Figure~\ref{fig:diff}(b), INDVE(minlog) performs within
seconds for the case of 100K variables and 100 to 6K ws-descriptors of
size $s = 2$, and with variable domain size $r = 4$. Two further
findings are not shown in the figure: (1) VE performs much worse than
INDVE, as it cannot exploit the independence of tuples and thus
creates partitions that overlap at large; (2) the case of $s=4$ has a
few (2 in 25) outliers exceeding 600 seconds.

\begin{figure}[t]
  \begin{center}
    \includegraphics[scale=1]{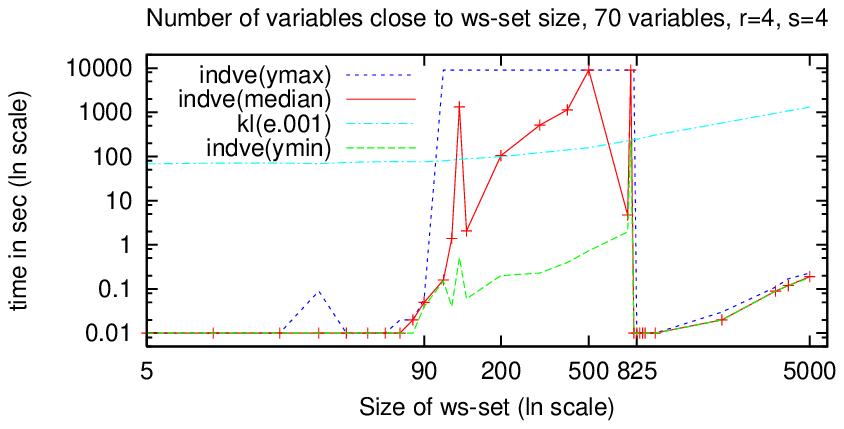}
  \end{center}
 \caption{Performance of INDVE and KL when numbers of variables and ws-descriptors are close.}
 \label{fig:smalldata}
 \vspace*{-1.5em}
\end{figure}

{\noindent\bf 3. The numbers of variables and of ws-descriptors are
close}. It is known from literature on knowledge compilation and model
counting~\cite{Birnbaum:DP:1999} that the computation becomes harder
in this case. Figure~\ref{fig:smalldata} shows the easy-hard-easy
pattern of INDVE(minlog) by plotting the minimal, maximal, and median
computation time of 20 runs (max allowed time of 9000s). We
experimentally observed the expected sharp transitions: When the
numbers of ws-descriptors and of variables become close, the
computation becomes hard and remains so until the number of
ws-descriptors becomes one order of magnitude larger than the number
of variables. The behavior of WE (not shown in the figure) follows
very closely the easy-hard transition of INDVE, but in our experiment
WE does not return anymore to the easy case within the range of ws-set sizes
reported on in the figure.

\begin{figure}[t]
  \begin{center}
    \includegraphics[scale=1]{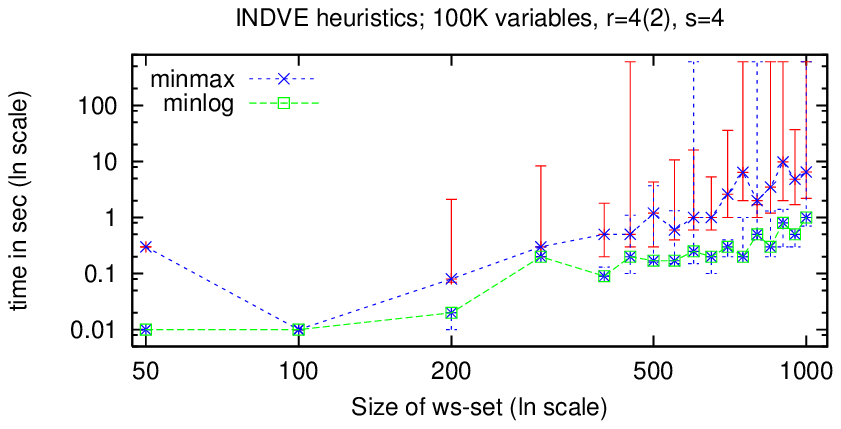}
  \end{center}
 \vspace*{-6mm}
 \caption{Heuristics: minmax versus minlog.}
 \label{fig:heuristics}
 \vspace*{-1.5em}
\end{figure}


{\noindent\bf 4. Exact versus approximate computation}.  We
experimentally verified our conjecture that the Karp-Luby
approximation algorithm (KL) converges rather slowly. In case the
numbers of variables and of ws-descriptors differ by orders of
magnitude, INDVE(minlog) and VE(minlog) are definitely competitive
when compared to KL with parameters 
$\epsilon = 0.1$ resp.\ $\epsilon=0.01$, and $\delta=0.01$, 
see Figure~\ref{fig:diff}.

In Figure~\ref{fig:diff}(b), KL uses about the same number of
iterations for all the ws-set sizes, a sufficient number to warrant
the running time. 
The reason for the near-constant line for KL is that for $s=2$ and
100k variables, ws-descriptors are predominantly pairwise independent,
and the confidence is close to $1-(3/4)^w$, where $w$ is the number of
ws-descriptors. But this quickly gets close to 1, and the optimal algorithm
can decide on a small number of iterations that does not increase with $w$.
In case the numbers of variables and 
ws-descriptors are close (Figure~\ref{fig:smalldata}), KL with
$\epsilon=0.001$ only performs better than INDVE(minlog) in the hard
cases.

{\noindent\bf 5. Heuristics for variable
elimination}. Figure~\ref{fig:heuristics} shows that, although the
minmax heuristic is cheaper to compute than the minlog heuristic,
using minlog we find in general better choices of variables and INDVE
remains less sensitive to data correlations. The plot data are
produced from 10 runs and show the median value and ymin/ymax for the
error bars. Although VE exceeds the allocated time of 600 seconds for
different data points, it does this less than five times (the median
value is closer to ymin).

\section{Related Work}
\label{sec:relatedwork}

To the best of our knowledge, this paper is the first to study the
conditioning problem for probabilistic databases. In this section, we
survey related work in the areas of probabilistic databases and
knowledge compilation procedures.

U-relations capture most other representation formalisms for uncertain
data that were recently proposed in the literature, including those of
MystiQ~\cite{dalvi07efficient}, Trio~\cite{BDSHW2006}, and
MayBMS~\cite{AJKO07}.  For each of these formalisms, natural
applications in data cleaning and other areas have been described
\cite{BDSHW2006,AKO07WSD,dalvi07efficient}.

Graphical models are a class of rich formalisms for representing
probabilistic information which perform well in scenarios in which
conditional probabilities and a known graph of dependencies and independences
between events are available. There are, for instance, Bayesian network
learning algorithms that produce just such data.
Unfortunately, if probabilistic
data is obtained by queries on tuple-independent or similar databases,
the corresponding graphical models tend to be relatively flat 
\cite{SD2007} but have
high tree-width, which causes techniques widely used for confidence
computation on graphical models to be highly inefficient.
%
%
Graphical models are more succinct than U-relations, yet
their succinctness does not benefit the currently known query evaluation
techniques. This justifies the development of conditioning techniques
specifically for the c-table-like representations (such as U-relations)
developed by the data\-base community.

It has been long known that 
computing tuple confidence values on DNF-like representations of sets of
possible worlds is a generalization of the DNF model counting problem
and is \#P-complete~\cite{DS2007}.
Monte Carlo approximation techniques for confidence computation have
been known since the original work by Karp, Luby, and Madras \cite{KLM1989}.
Within the database field, this approach has first been followed in
work on query reliability \cite{GGH1998} and in
the MystiQ project \cite{dalvi07efficient}.
Section~\ref{sec:experiments} reports on an experimental
comparison of approximation and our exact algorithms.

Our variable elimination technique is based on Davis-Put\-nam
procedure for satisfiability checking~\cite{DP1960}. This procedure
was already used for model counting~\cite{Birnbaum:DP:1999}. Our
approach combines it with independent partitioning for efficiently
solving two more difficult problems: exact confidence computation and
conditioning. \cite{Birnbaum:DP:1999} uses the minmax heuristic (which
we benchmark against) and discusses experiments for CNF formulas with
up to 50 variables and 200 clauses only. Our experiments also discuss
new settings that are more natural in a database context: for
instance, when the size of a query answer (and thus the number of
ws-descriptors) is small in comparison to the size of the input
database (and thus of variables). Follow-up work~\cite{HO:SUM:2008}
reports on techniques for compiling ws-sets generated by conjunctive
queries with inequalities into decision diagrams with polynomial-time
guarantees.

Finally, there is a strong connection between ws-trees and ordered
binary decision diagrams (OBDDs). Both make the structure of the
propositional formulas explicit and allow for efficient manipulation.
\nop{The most studied decision diagrams are the ordered binary decision
diagrams (OBDDs)~\cite{Meinel:OBDD:1998}. In short, an OBDD is a
directed acyclic graph with inner nodes labeled by Boolean variables
and two leaves annotated by \textit{true} and \textit{false},
respectively. Each node has two outgoing edges corresponding to the
two possible Boolean values of its variables. All root-to-leaf paths
have the same variable ordering. OBDDs have important properties:
checking satisfiability and equivalence is in constant time (all
equivalent formulas have one canonical OBDD) and counting the number
of satisfying assignments (or models) can be done in linear time. Note
that model counting is a special instance of probability computation,
where we assume that all satisfying assignments (worlds) have the same
weight (probability).}
They differ, however, in important aspects: binary versus multistate
variables, same variable ordering on all paths in case of OBDDs, and
the new ws-tree $\otimes$-node type, which makes independence
explicit. It is possible to reduce the gap between the two formalisms,
but this affects the representation size. For instance, different
variable orderings on different paths allows for exponentially more
succinct BDDs~\cite{Meinel:OBDD:1998}. Multistate variables can be
easily translated into binary variables at a price of a logarithmic
increase in the number of variables~\cite{Wachter:Multistate:2007}.

\nop{
The $\otimes$-nodes could be prohibited in ws-trees at the price of an
exponential increase in the representation
size. Figure~\ref{fig:ws-tree-elim} gives a translation of ws-trees
into free decision diagrams with multistate variables by eliminating
the $\otimes$-nodes. A ws-tree rooted at a $\otimes$-node with
$S_1,\ldots,S_{|I|}$ children is translated into a ws-tree rooted at
$S_1$ and where all leaves of $S_1$ are replaced by $S_2$, all leaves
of $S_2$ are replaced by $S_3$, and so on. This translation can lead
to exponentially larger decision diagrams. We can obtain a translation
to linear-size free decision diagrams (which are directed
acyclic graphs) by keeping one copy of each subtree $S_i$ and have
links to that copy instead of replicating $S_i$.

\begin{figure}
\framebox[\columnwidth]{
\hspace{2mm}
\parbox{8cm}{
\begin{align*}
f \big(\bigotimes_{i \in I} S_i\big) & = f (\mu (S_1,
[S_2,\ldots,S_{|I|}]))\\ f \big( \bigoplus_{i \in I} (x \mapsto i :
S_i) \big) &= \bigoplus_{i \in I} (x \mapsto i : f(S_i))\\ 
f(\emptyset) &= \emptyset \hspace*{4em} f(\bot) = \bot\\
\mu \big(\bigotimes_{i \in I} S_i, h::t\big) &= \bigotimes_{i \in I} (\mu(S_i, h::t))\\
\mu \big(\bigoplus_{i \in I} (x \mapsto i : S_i), h::t\big) &= \bigoplus_{i \in I} (x \mapsto i: \mu(S_i, h::t))\\
\mu (\emptyset, h::t) &= \mu (h,t)\\
\mu (S,[]) &= S
\end{align*}
}}

\caption{Elimination of $\otimes$-nodes in ws-trees.}
\label{fig:ws-tree-elim}
\end{figure}
}

\newpage

\bibliographystyle{abbrv}
\begin{small}

\end{small}

\newpage
\appendix

\subsection*{Proof of Theorem~\ref{th:ws-tree}}

We prove that the translation from ws-sets to ws-trees is correct.
That is, given a ws-set $S$, \mbox{ComputeTree}($S$) and $S$ represent
the same world-set.

We use induction on the structure of ws-trees. In the base case, we
map ws-sets representing the empty world-set to $\bot$, and ws-sets
containing the universal ws-descriptor $\emptyset$ (that represents
the whole world-set) to $\emptyset$. We consider now a ws-set $S$. We
have two cases corresponding to the different types of ws-tree inner
nodes.

Case 1.  Assume $S=\underset{i\in I}\bigcup S_i$ with $S_i$ pairwise
independent and $R_i=\mbox{ComputeTree}(S_i)$. By hypothesis, $\omega(R_i) =
\omega(S_i)$. Then, $\mbox{ComputeTree}(S) = \underset{i\in I}{\bigotimes} (R_i)$ and
$\omega(\mbox{ComputeTree}(S)) = \underset{i\in I}{\bigcup}
\omega(R_i) = \underset{i\in I}{\bigcup} \omega(S_i) = \omega(S).$

Case 2. Let $x$ be a variable in $S$ and consider the ws-sets
$S_{x\mapsto i}$ ($i\in\mbox{dom}_x$) and $T$ as given by ComputeTree.
Because the whole world-set can be represented by $A =
\underset{i\in\mbox{dom}_x}{\bigcup} \{\{x\mapsto i\}\}$, it holds
that $\omega(A)\cap\omega(S) = \omega(S)$. We push the assignments of
$x$ in each ws-descriptor of $S$ and obtain $$\omega(S) =
\omega\big(\underset{i\in\mbox{dom}_x}{\bigcup} \{d\cup\{x\mapsto
i\}\mid d\in S\}\big).$$
We can remove all inconsistent ws-descriptors in the ws-set of the
right-hand side while preserving equivalence:
\begin{align*}
&\omega(\{d\cup\{x\mapsto i\}\mid d\in S\}) =\\
&\omega(\{d\cup\{x\mapsto i\}\mid d\in S,\not\exists j\in\mbox{dom}_x: j\not=i, \{x\mapsto j\}\subseteq d\})=\\
&\omega(\{d\cup\{x\mapsto i\}\mid \{x\mapsto i\}\subseteq d\in S\})\cup\\
&\omega(\{d\cup\{x\mapsto i\}\mid d\in S, \not\exists j\in\mbox{dom}_x: \{x\mapsto j\}\subseteq d\})=\\
&\omega(S_{x\mapsto i})\cup\omega(T) = \omega(S_{x\mapsto i}\cup T)
\end{align*}
We now consider all values $i\in\mbox{dom}_x$ and obtain
\begin{align*}
\omega(S) &=\omega\big(\underset{i\in\mbox{ dom}_x}{\bigcup} (S_{x\mapsto i}\cup T)\big)\\ 
&=\omega\big(\underset{i\in\mbox{ dom}_x}{\bigoplus} x\mapsto i: (S_{x\mapsto i}\cup T)\big).
\end{align*}

\subsection*{Proof of Theorem~\ref{th:conditioning}}

We prove that given a representation of probabilistic da\-ta\-base
${\bf W} = \{ (I_1, p_1), \dots, (I_n, p_n) \}$ and a ws-tree $R$
identifying a nonempty subset of the worlds of {\bf W}, the algorithm
of Figure~\ref{fig:conditioning_algorithm} computes a representation
of probabilistic database
\[
\{ (I_j, \frac{p_j}{c}) \mid (I_j, p_j) \in {\bf W}, I_j \in \omega(R) \} 
\]
such that the probabilities $p_j$ add up to 1.

The conditioning algorithm computes the probability $c$ of each node
of the input ws-tree $R$ as given by our probability computation
algorithm of Figure~\ref{fig:ws-tree-prob}.  We next consider the
correctness of renormalization using induction on the structure of the
input ws-tree.

Base case: The ws-tree $\emptyset$ represents the whole world-set and
we thus return $U$ unchanged (no conditioning is done).

Induction cases $\otimes$ (independent partitioning) and $\oplus$
(variable elimination). For both node types, we return the union of
ws-sets $U'_i$ that are the ws-sets $U_i\subseteq U$ where the
variables encountered at the nodes on the recursion path are replaced
by new ones. The ws-sets $U_i$ are the subsets of $U$ consistent with
each child of the $\oplus$ or $\otimes$ node. By hypothesis, the
ws-sets $U_i$ are conditioned correctly. In case of $\otimes$-nodes,
no further conditioning is done, because no re-weighting takes
place. In case of a $\otimes$-node, we re-weight the assignments of
the variable eliminated at that node.

Let $I\subseteq \mbox{dom}_x$ be the set of alternatives of $x$
present at that node. Since
\[
P(R) = P \big( \bigoplus_{i \in I} (x \mapsto i : R_i) \big)
       = \sum_{i \in I} P(\{x \mapsto i\}) \cdot P(R_i),
\]
if we create a new variable $x'$,
\[
P(\{x' \mapsto i\}) := \frac{P(\{x \mapsto i\}) \cdot P(R_i)}{P(R)}.
\]
This guarantees that
\[
P \big( \bigoplus_{i \in I} (x' \mapsto i : R_i) \big) = 1.
\]

If we ask which tuples of $U$ should be in an instance satisfying $R$,
the answer is of course all those whose ws-descriptors are consistent with
one of the ws-descriptors in $x \mapsto i : R_i$ for some $i \in I$.
The $U$-relation tuples in the results of the invocations cond($R_i$, $U_i$)
grant exactly this.

\end{document}